\documentclass[aps,pre,preprint,superscriptaddress,showkeys]{revtex4-1}

\usepackage{graphicx}
\usepackage{amsmath}
\usepackage{verbatim}
\usepackage{textcomp}
\usepackage{color}
\graphicspath{{./pics/}}

\begin{document}


\title{Self-assembly of ``Mickey Mouse'' shaped colloids into tube-like structures: experiments and simulations}


\author{Joost R Wolters}
\email[]{j.r.wolters@uu.nl}
\affiliation{Utrecht University, Debye Institute for Nanomaterials Science}

\author{Guido Avvisati}
\affiliation{Utrecht University, Debye Institute for Nanomaterials Science}

\author{Fabian Hagemans}
\affiliation{Utrecht University, Debye Institute for Nanomaterials Science}

\author{Teun Vissers}
\affiliation{University of Edinburgh}

\author{Daniela J. Kraft}
\affiliation{Leiden University, Soft Matter Group}

\author{Marjolein Dijkstra}
\affiliation{Utrecht University, Debye Institute for Nanomaterials Science}

\author{Willem K. Kegel}
\email[]{w.k.kegel@uu.nl}
\affiliation{Utrecht University, Debye Institute for Nanomaterials Science}


\date{\today}

\begin{abstract}
The self-assembly of anisotropic patchy particles with triangular shape was studied by experiments and computer simulations. The colloidal particles were synthesized in a two-step seeded emulsion polymerization process, and consist of a central smooth lobe connected to two rough lobes at an angle of $\sim$90\textdegree , resembling the shape of a ```Mickey Mouse'' head. Due to the difference in overlap volume, adding an appropriate depletant induces an attractive interaction between the smooth lobes of the colloids only, while the two rough lobes act as steric constraints.
The essentially planar geometry of the ``Mickey Mouse'' particles is a first geometric deviation of dumbbell shaped patchy particles. This new geometry is expected to form one-dimensional tube-like structures rather than spherical, essentially zero-dimensional micelles.
At sufficiently strong attractions, we indeed find tube-like structures with the sticky lobes at the core and the non-sticky lobes pointing out as steric constraints that limit the growth to one direction, providing the tubes with a well-defined diameter but variable length both in experiments and simulations. In the simulations, we found that the internal structure of the tubular fragments could either be straight or twisted into so-called Bernal spirals.
\end{abstract}


\maketitle


  \section{
      \label{sec:intro}Introduction}
      The geometry of a self-assembled object is expected to depend on the shape and specific interactions of the individual building blocks~\cite{sacanna2013shaping, yi2013recent}. This is seen for instance in systems of surfactants, where the geometry of the surfactant molecules determines the geometry of the micelles they form~\cite{Israelachvili1976}. Similarly, conical dendrons are found to form spherical supramolecular dendrimers, whereas flat, tapered dendrons form cylindrical dendrimers~\cite{Percec2000}.
By changing the shape-anisotropy of the colloidal building blocks with attractive and non-attractive patches, we can tune the directionality of the interactions for colloidal self-assembly. In this paper, we introduce new colloidal building blocks consisting of one sticky and two non-sticky lobes in a triangular conformation. Since the second non-sticky lobe provides our particles with a planar rather than conical geometry, we expect these particles to induce the formation of cylindrical aggregates.
          
There are several ways to introduce specific interactions to colloidal particles. Spherical particles with a single attractive patch, also called Janus particles, can be made by templated surface modification~\cite{Hong2008,Jiang2010} and have been found to form small clusters and chain-like aggregates~\cite{Hong2008, iwashita2013stable}. This synthesis method can be extended to multiblock particles, introducing additional attractive patches~\cite{Chen2012}, allowing for the formation of for instance Kagome lattices in 2D upon sedimentation~\cite{Chen2011a}. Another way to introduce attractive patches to spherical colloids is by making use of complementary strands of DNA. This method was employed by Pine and co-workers~\cite{Wang2012} in a multi-step synthesis method to create particles with up to four well-defined sticky patches, mimicking molecular bonds and forming the colloidal equivalent of molecules. Since these systems are based on spherical particles, self-assembly is based on the topology of attractive patches on the particles and not on the particle shape.

In addition to changing the topology of attractive patches on the surface, interaction can also be made directional by altering the shape of the particle such that it becomes non-spherical. A good example of this shape-anisotropy are dumbbell-shaped colloids that can be used as building blocks for colloidal self-assembly. Such particles can be produced in bulk by forming a protrusion on a polymer seed particle using seeded emulsion polymerization~\cite{Sheu1990,Mock2006,Kim2006,Park2010,Nagao2011,Kim2008}
and if the two sides of such a particle are chemically different~\cite{Kim2006,Kim2008,Nagao2011,VanRavensteijn2013}, this difference can be used to induce specific interactions to either lobe of the dumbbell.

Because it is easier to tune the shape and interaction of particles in simulations, a lot of exploratory simulation work on the self-assembly behavior of patchy dumbbell and Janus particles has already preceded the experiments, finding finite sized micelle-like clusters~\cite{Chen2007,Chen2007a, munao2013structure, munao2014phase} and vescicles, tubular, lamellar and crystalline equilibrium phases~\cite{Sciortino2009, Sciortino2010, januscrystals, munaotubes, Preisler2013, coop}. However, even if such particles can be synthesized, their specific interactions are usually short-ranged and strong, so not all structures predicted by simulations can be achieved experimentally.

Recently, Kraft \emph{et al.}~\cite{Kraft2012} introduced specific interactions in a system of asymmetric dumbbells by site-specific surface roughness. Upon addition of a depletant, only the smaller smooth lobe of the particle became attractive, which resulted in finite sized, spherical micelle-like clusters.

In this paper, we extended these experiments to particles with a new shape by introducing an additional rough lobe. These ``Mickey Mouse'' (MM) shaped particles consist of a smooth lobe that becomes attractive in the presence of a depletant and two rough lobes, its ``ears'', at an angle close to 90\textdegree~ that remain non-attractive (Fig.~\ref{fig:dimers}). 
These two non-attractive lobes provide the particles with a planar geometry and give directionality to the attractions between the smooth lobes in such a way that they are expected to form long, one-dimensional tube-like structures that rarely branch.

Synthesis of MM particles is achieved by forming a single smooth protrusion on small clusters of rough seed particles. Subsequently, the desired particles, a smooth protrusion on a dimer of rough particles, can be isolated in sufficient yield by density gradient centrifugation. The effective attraction between the smooth parts of the MM particles is tuned by variation of the depletant concentration. To investigate the effect of particle concentration and interaction strength on the resulting structures, we use a combination of optical microscopy and Monte Carlo computer simulations. From this combined approach, it becomes clear that the unique geometry of the particles indeed causes the formation of elongated tube-like structures with a well-defined diameter. However, this particle geometry and the short-ranged nature of the interaction also seem to prevent the system from converging to its ground state.

\section{\label{sec:experimental}Experimental}

\subsection{Outline of the Synthesis Procedure}
The MM particles used in this experiment were synthesized in three steps. The initial step is a dispersion polymerization. The two subsequent steps are seeded emulsion polymerization procedures, using the product of the previous step as seeds. These two steps are a modified version of the synthesis described by Kim \emph{et al.}~\cite{Kim2006}, and later also used by Kraft \emph{et al.}~\cite{Kraft2012}. In the first step, spherical, linear polystyrene particles were synthesized, using a charge free initiator and polyvinylpyrrolidone (PVP) as steric stabilizer. In the second step, spherical, cross-linked polystyrene particles with a rough surface were formed by swelling the seed particles with monomer and subsequent polymerization. The roughness consisted of smaller particles, formed by secondary nucleation during the reaction, which attached to the particles. In this synthesis, this step did not only produce single rough particles, but also a significant amount of small clusters of 2, 3, 4 or more of these rough particles. In the third step, the dispersion of rough particles with this cluster distribution was swollen again with monomer, now forming a liquid protrusion on the surface of the seed particles. Due to the presence of these n-mers in the seed dispersion, this step did not only produce dumbbells, but also clusters of n rough particles sharing a single smooth protrusion.

After polymerization, the clusters of different sizes were separated using density gradient centrifugation (DGC) and the clusters of two rough particles with one protrusion were isolated. The self-assembly of these MM particles is reported here, the other particle shapes formed in this synthesis will be investigated in other experiments.

\subsection{Particle Synthesis}
For the dispersion polymerization, styrene (Sigma-Aldrich, Reagent Plus
) was used as received. For the other two steps, it was distilled to remove the inhibitor. All other chemicals were used as received. 

In the first step, 10~mL styrene, 5.0~g PVP (Sigma-Aldrich, $M_{w}$ = 40~kgmol$^{-1}$) and 0.136~g azobisiobutyronitrile (AIBN, Sigma-Aldrich) were dissolved in 126~mL ethanol (p.a., Merck) and 14~mL H$_{2}$O (Milli\-pore
) in a 250~mL round bottom flask. The flask was flushed with N$_{2}$ and sealed tightly with a stopper and Teflon tape. Polymerization was carried out by immersing the flask in an oil bath of 70~\textcelsius{} at an angle of 60\textdegree~ and rotating it at $\sim$100~rpm for 24 hours. The resulting dispersion was washed by centrifugation and redispersion in methanol three times to remove remaining reactant and transferred to H$_{2}$O for further processing.

For the second step, the cross-linking of the particles, a monomer solution was prepared, consisting of styrene (distilled), 9\%~v/v 3-(Tri\-methoxy\-silyl)propyl\-methacrylate (TPM, 98\% Acros) as a co-monomer, 
 1.3\%~v/v divinylbenzene (DVB, 80\% tech. Aldrich) as a cross-linker and 2\%~wt V-65 (Wako chemicals GmbH) as an initiator.

A swelling emulsion was prepared by adding this apolar monomer solution to an water phase containing 1.3\%~wt polyvinyl-alcohol (PVA, $M_{w}$ = 85-124~kgmol$^{-1}$, 87-89\% hydrolysed, Sigma-Aldrich) to stabilize the emulsion. A total of 10~mL in a monomer:water ratio of 1:4 was added to an elongated 25~mL vial. The content of the vial was then emulsified for 5 minutes at 8~krpm using an IKA T-25 Ultra Turrax with an S25N 10G dispersing tool.

This swelling emulsion was added to a 20\%~v/v dispersion of the seed particles from the first step, so that the swelling ratio $S = m_{monomer}/m_{polymer seeds}$ equals 4. This vial was flushed with N$_{2}$ for $\sim$30~s, sealed with Parafilm and put on a RM5-80V tube roller (50~rpm) for 18 hours to allow for the particles to take up the monomer. After this, polymerization is carried out by immersing the vial in a 70~\textcelsius~ oil bath at an angle of 60\textdegree~ and rotating it at  $\sim$100~rpm for 24 hours. After polymerization, the particles were washed several times by centrifugation and redispersion in H$_{2}$O to get rid of any secondary nucleation not attached to the particles.

In the third step a protrusion is formed on the now cross-linked particles. For this, the procedure of emulsification, swelling and polymerization was essentially repeated, albeit with a few differences: this time, the monomer solution did not contain TPM as co-monomer and the emulsion contained only half the amount of monomer solution (1:9 monomer:water), to achieve a swelling ratio of $S = 2$. Furthermore, the PVA concentration in the aqueous phase was only 0.5\%~wt instead of the 1.3\%~wt used in the second step. After polymerization, the particles were again washed several times by centrifugation and redispersion in H$_{2}$O until the supernatant was clear.

\subsection{Density Gradient Centrifugation}
The MM particles were isolated from the mixture by density gradient centrifugation (DGC). Using a Gilson Minipuls 3 peristaltic pump (5 rpm), an Ultra-Clear centrifuge tube was carefully filled with 30~mL of a density gradient ranging from 9\% to 3\%~wt Ficoll PM~400 (Sigma-Aldrich) in an aqueous solution of 1\%~wt Pluronic F-127 (Sigma-Aldrich).
1~mL of a 1\%~v/v dispersion of the mixed particles in H$_{2}$O was subsequently placed on top of the density gradient and the tube was centrifuged for 20 minutes at 1000~rpm (233 $\times g$).

The second band, containing the desired particles, was then extracted by penetrating the tube wall with a needle connected to the peristaltic pump. This dispersion was concentrated by centrifugation and redispersion in an aqueous solution of 0.3\%~wt PVA ($M_{w}$ = 31-50~kgmol$^{-1}$, 87-89\% hydrolyzed, Sigma-Aldrich) to a particle concentration of $\sim$3\%~v/v.

\subsection{Sample Preparation}
In order to study the self-assembly of MM particles, samples with varying particle and depletant (dextran) concentrations were prepared. Capillaries (0.10~mm~$\times$~2.00~mm internal dimensions, Vitrotubes W5010-050) were filled with these samples. In order to prevent interaction with the capillary wall, these capillaries were first coated with dextran~\cite{Badaire2007}.

To coat the capillaries, a sequence of solutions was pumped through them, again using a peristaltic pump. Consecutively, 0.5~mL each of 1~M NaOH (to negatively charge the capillary wall), 1\%~wt polyethyleneimine (PEI, $M_{w}$ = 60~kgmol$^{-1}$, 50\% in H$_{2}$O, Sigma-Aldrich) (a cationic polymer) and 1\%~wt dextran sodium sulfate (DSS, $M_{w}$ = 500~kgmol$^{-1}$, Fisher Scientific) (the same as the depletant, but carrying a negative charge) were flushed through the capillaries at a pump speed of 1.20~rpm, each solution followed by a rinsing step with 0.5~mL H$_{2}$O. After this, the capillaries were left submersed in H$_{2}$O for 30 minutes and dried under a flow of N$_{2}$.

Sample mixtures were made with different particle and depletant concentrations. Typically, 20~\textmu L of particle dispersion of a known concentration from DGC is mixed with  2.5 to 5.0~\textmu L of 116~gL$^{-1}$ dextran solution (from Leuconostoc spp., $\sim$500~kgmol$^{-1}$, Sigma-Aldrich), 2.5 to 3.0~\textmu L 77~mM sodium azide (99\% extra pure, Merck), 1~\textmu L 1~M NaCl and 23.45~\textmu L D$_{2}$O. This results in samples with a particle volume faction $\phi_{\mathrm{particles}}$ of 0.003 or 0.01, containing 24~mM of salt and a depletant volume fraction $\phi_{d}$ of 0.20 to 0.40 times the overlap volume fraction. The coated capillaries were filled with depletion sample and glued to object slides using UV-curable glue (Norland Optical Adhesive 81).
D$_{2}$O was added to reduce the density difference between particles and medium, reducing sedimentation. Nevertheless, in the capillaries particles and clusters still sedimented within a day, so they had to be stored on a RM5-80V tube roller (50~rpm) to keep them suspended.

\subsection{Analysis}
Particle size and shape were determined by analysing TEM images 
 taken 
using a FEI Tecnai 10 transmission electron microscope. The particle surface roughness was investigated using scanning electron microscopy (SEM XL FEG 30, Philips). 
A Malvern ZetaSizer ZS was used to determine polymer size using Dynamic Light Scattering (DLS) and particle zeta potential using Laser Doppler Electrophoresis. Optical microscopy investigation of self-assembly was carried out using a Zeiss Axioplan optical microscope using an oil-immersion objective (NA = 1.4, 100$\times$ magnification). Images were captured using a Basler Scout camera and the Streampix software package.

The free particle concentration in the samples was determined by making scans at several places in the capillary, imaging the entire capillary in the $z$-direction (100~\textmu m) and counting the number of free particles in the sample volume. This was translated to a volume fraction using the volume of one particle as calculated from the TEM data ($v_{\mathrm{MM}} = 21$~\textmu m$^{3}$). Error bars were calculated by taking the standard deviation in particle concentrations obtained from different scans.

       \subsection{Simulation Model}
In the computer simulations, only the essential features of the experimental particles were modelled. An MM particle is represented by an aggregate of three spheres, one central sphere with diameter $\sigma_{\mathrm{s}}$ which represents the smooth lobe, and on either side of it two smaller spheres with diameter $\sigma_{\mathrm{r}}$ mimicking the rough lobes. The attraction between two smooth lobes $i$ and $j$ of two different MM particles is described by a square-well interaction,
\begin{equation}
        u^\mathrm{SW}(r_{ij})=\left\{
                        \begin{array}{lll}
                                \infty & \mbox{if } &r_{ij} < \sigma_{\mathrm{s}}\\
                                \varepsilon & \mbox{if } &\sigma < r_{ij} < \sigma_{\mathrm{s}}+\Delta\\
                                                                0 & \mbox{if } &r_{ij}>\sigma_{\mathrm{s}}+\Delta\\
                        \end{array}
                        \right.,
                        \label{int:sw}
\end{equation}
where $\varepsilon < 0$
 denotes the depth of the well, and $\Delta$ is the range of the interaction. The rough lobes of the particles were treated as non-overlapping hard bodies, where the interaction between two rough lobes $k$ and $m$ of two different particles is described with a hard-sphere potential, 
\begin{equation}
        u^\mathrm{HS}(r_{km})=\left\{
                        \begin{array}{lll}
                                \infty & \mbox{if } &r_{km} < \sigma_{\mathrm{r}}\\
                                0 & \mbox{if } &r_{km} >\sigma_{\mathrm{r}}\\
                        \end{array}
                        \right..
\end{equation}
Note that, in spite of using a spherically symmetric potential for the attractions, the total particle-particle interaction is not invariant under rotations, due to the steric constraint of the rough lobes.
          
All Monte Carlo (MC) simulations were performed in the canonical ensemble (MC-NVT) of $N=256$ MM particles in a volume $V$ and a temperature $T$, employing single particle translations and rotations~\cite{bib:Frenkel.ums} to explore the configurational phase space. The simulations were performed for $1-4 \times10^8$ MC steps, where a single Monte Carlo step is defined as $N$ attempted moves (either translations or rotations). The effect of varying the packing fraction $\phi=\rho v_{\mathrm{MM}}$ (where $\rho=N/V$ and $v_{\mathrm{MM}}$ is the volume of a single particle) and the interaction energy $\beta\varepsilon$ (Here $\beta = 1/k_{\mathrm{B}}T$ denotes the inverse temperature with $k_{\mathrm{B}}$ Boltzmann's constant and $T$ the temperature) was investigated, while the other parameters in the model were fixed to match the experimental system as closely as possible.
          
The simulation results were analyzed using a cluster analysis. To this end, a cluster criterion was employed to distinguish the clusters in the simulations. For this, MM particles were identified as interacting neighbors that have at least one mutual bond, i.e. the distance between their attractive lobes is less than the square-well interaction range. Subsequently, a cluster was defined as a contiguous set of neighboring particles. The cluster analysis allows us to compute for each interaction strength the concentration of free particles in the system, as well as the number of bonds per particle in the clusters.
          
\section{\label{sec:results}Results and Discussion}

\subsection{Particle Synthesis}
The dispersion polymerization of seed particles yielded spherical polystyrene particles with a radius of $0.617 \pm 0.011$~\textmu m (1.7\% polydispersity). After the first seeded emulsion polymerization step, particles had a radius of $0.86 \pm 0.03$~\textmu m (4\% polydispersity) and a surface covered with small particles formed as a result of secondary nucleation. Furthermore, optical microscopy showed the sample contained single particles ($\sim$60\%), but also dimers ($\sim$30\%) and trimers and larger clusters ($\sim$10\%).

Using these particles in the protrusion formation yielded a mixture of different sized clusters with a single smooth protrusion. The dispersion contained single dumbbells, dimers of two rough particles sharing one smooth protrusion, and multimers of more than two particles sharing a protrusion. The ratios in which these different particle shapes occurred had the same distribution as the seed particles used for the reaction.

It is interesting to note that while the result is similar to the colloidal molecules produced previously by Kraft \emph{et al.}~\cite{Kraft2009a,Kraft2009}, the mechanism by which they are formed is different. Since the distribution of clusters in the seed dispersion was found to be the same as the distribution of colloidal molecules formed (not shown),  the colloidal molecules must be formed from clusters already present in the seed dispersion, not by fusion of the liquid protrusions during the third step.

After separation of the different fractions by DGC, the MM particles were successfully isolated. Although the separation using DGC is not perfect, analysis of the SEM images shows that all MM particles are uniform in shape and the fraction of undesired particles is very small. These impurities are therefore not considered to play a role in the structures formed by self-assembly of the MM particles. In order to acquire sufficient particles for all samples, several parallel DGC steps were carried out.
\begin{figure}[h]
\centering

        \includegraphics[width=0.5\textwidth]{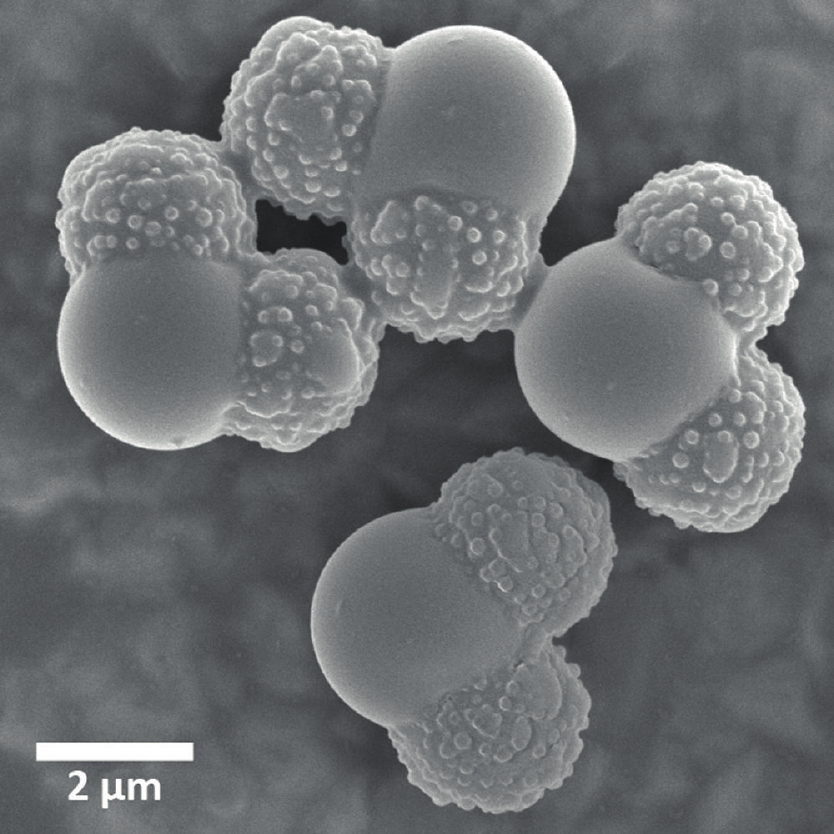}

\caption{\small{SEM micrograph of the isolated MM particles\label{fig:dimers}}}
\end{figure}
As is shown in Fig.~\ref{fig:dimers}, the particles have one smooth lobe ($R_{s} = 1.38 \pm 0.10$~\textmu m, 7\% polydispersity) and two lobes made rough by irreversibly attached secondary nucleation ($R_{r} = 1.17 \pm 0.07$~\textmu m, 6\% polydispersity). The angle between the lobes with respect to the center of the smooth lobe is $91^{\circ}$, with some variation due to the polydispersity of the particles.

While the particles rely on PVA on their surface for steric stabilization, their surface charge also contributes significantly to their interaction potential. Therefore, the zeta potential of the particles was determined. From their electrophoretic mobility, the particle zeta potential was found to be  $-32 \pm 4$~mV, assuming spherical particles with uniform charge distribution. 
Since our particles are not spherical and the charged groups are expected to be mostly present on the rough lobes, this value is only regarded as an estimate of the charge on the smooth lobes that ultimately contributes to the net attractive potential.

To match the experimentally measured properties of the MM particles, the particles in the simulation were chosen to have a size ratio $q=\sigma_{\mathrm{r}}/\sigma_{\mathrm{s}}=0.85$, a center-to-center distance between the smooth and rough lobes of $d=0.57\sigma_{\mathrm{s}}$ and an angle between the rough lobes of $91^{\circ}$. The range of interaction $\Delta=0.02 \sigma_{\mathrm{s}}$, where $\sigma_{\mathrm{r}}$ and $\sigma_{\mathrm{s}}$ are the diameters of the rough and smooth lobes respectively. 

\subsection{Self-assembly}
\subsubsection{Interaction potential in experiments}
In the experiments, the attractive interactions between the smooth lobes of the particles as function of their center-to-center separation $r_{ij}$ results from an attractive depletion interaction and a repulsive electrostatic interaction:
\begin{equation}
w(r_{ij})=w_{el}(r_{ij})+w_{depl}(r_{ij})
\label{eq:netinteraction}
\end{equation}
Here, the electrostatic repulsion (in units of $k_{\text{B}}T$) can be described as~\cite{Verwey1948}:
\begin{equation}
w_{el}(r_{ij})/k_{\mathrm{B}}T = \frac{(\Psi/k_{\mathrm{B}}T)^{2}R_{s}^{2}}{\lambda_{B}r_{ij}}e^{-\kappa(r_{ij}-2R_{s})}
\label{eq:wel}
\end{equation} 
where $\Psi = e\zeta$ is the zeta potential times the elementary charge~\cite{Hsu2005}, $R_{s}$ is the radius of the smooth lobe of the particle, $\lambda_{B}$ is the Bjerrum 
length and $\kappa$ is the inverse Debye length.

The depletion attraction between two-spheres, for $ 2R_{s} < r_{ij} < 2(R_{s}+r_{g})$ can be written as~\cite{Asakura1958,Vrij1976,Lekkerkerker1992}:
\begin{equation}
w_{depl}(r_{ij})/k_{\mathrm{B}}T = -\phi_{d}\left(\frac{R_{s}+r_{g}}{r_{g}}\right)^{3}\left(1-\frac{3}{4}\frac{r_{ij}}{R_{s}+r_{g}}+\frac{1}{16}\left(\frac{r_{ij}}{R_{s}+r_{g}}\right)^{3}\right) 
\label{eq:wdepl}
\end{equation}
where $r_{g}$ is the radius of gyration of the depletant polymer and $\phi_{d}$ is the depletant volume fraction, as fraction of the overlap concentration.
\begin{figure}[h]
\centering

        \includegraphics[width=0.5\textwidth]{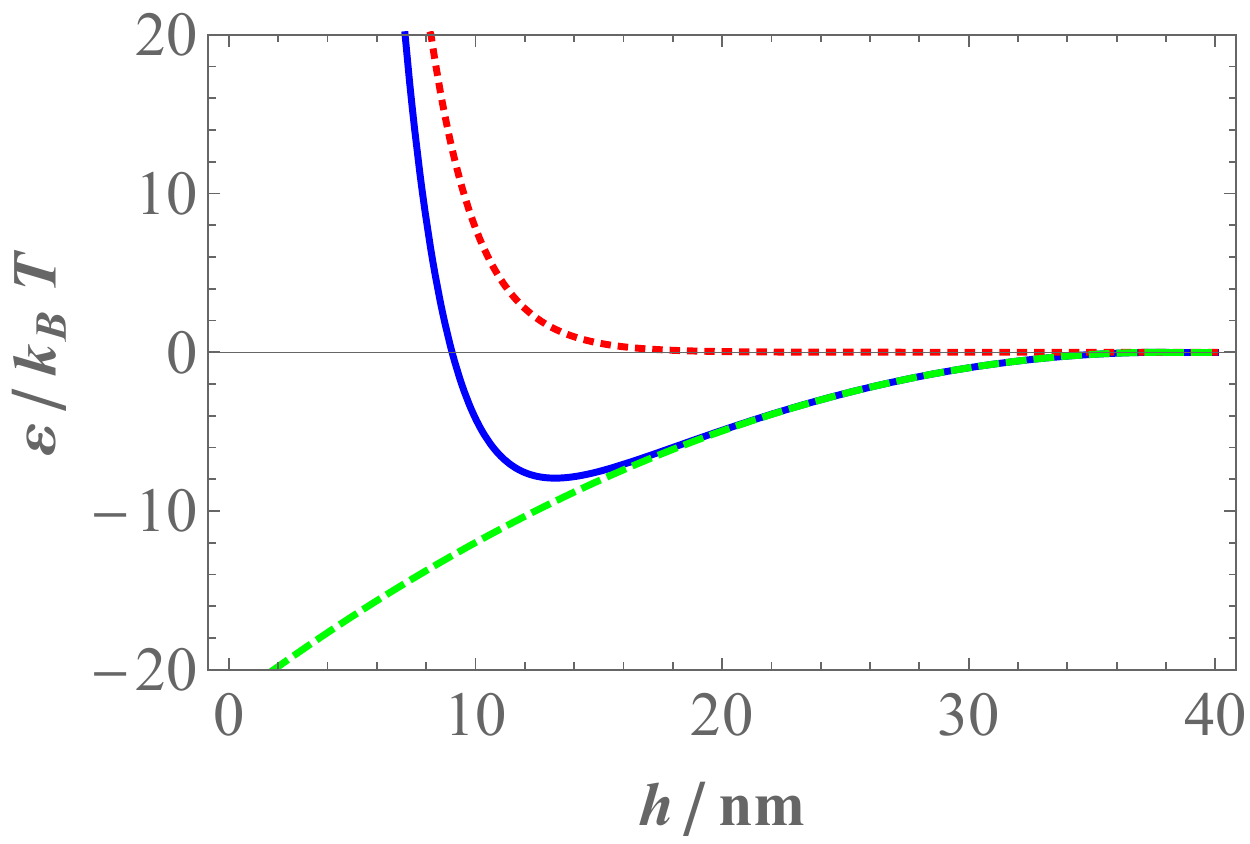}

\caption{\small{Interaction potential in $k_{\text{B}}T$ between the smooth lobes of the particles as function of the inter particle distance $h=r_{ij}-2R_{s}$ without considering the presence or orientation of the rough lobes. The net interaction (Eq.~\ref{eq:netinteraction}, solid blue line) is the sum of the screened coulomb repulsion (Eq.~\ref{eq:wel}, red dotted line) and the depletion attraction (Eq.~\ref{eq:wdepl}, dashed green line). Typical experimental parameters were used to construct this graph, resulting in a depth of the potential well $\varepsilon$ of about $-8\;k_{\text{B}}T$ and an interaction range of 40~nm.
\label{fig:pairpotential}}}
\end{figure}
Using experimental values for $R_{s}$, $\phi_{d}$, $r_{g}$ (19~nm, determined by dynamic light scattering), the salt concentration ($I(M)=25$~mM, $\kappa=0.50$~nm$^{-1}$), and $\Psi$ in Eq.~\ref{eq:wel} and \ref{eq:wdepl}, we find that the minimum of the overall potential varies in the range of $-7\;k_{\text{B}}T$ to $-17\;k_{\text{B}}T$ for a depletant volume fraction $\phi_{d}$ of 0.20 to 0.40 (The range of experimentally used values). Note that a zeta potential of $-32$~mV is used to find $\Psi=1.2\;k_{\text{B}}T$.
The square-well attractions (Eq.~\ref{int:sw}) in the simulations and the experimental potential (Eq.~\ref{eq:netinteraction} and Fig.~\ref{fig:pairpotential}), have a different shape, but a comparable range ($\Delta=0.02 \sigma_{\mathrm{s}}$ corresponds to $55$~nm in the experimental system) which is small with respect to the particle diameter. 
According to the Noro-Frenkel law of corresponding states \cite{noro2000extended, foffi2007possibility}, short-ranged potentials give similar phase behavior if their second virial coefficients are similar. Using Eq.~\ref{eq:netinteraction} for the experimental potential and Eq.~\ref{int:sw} for the square-well attractions in the simulations gives similar values of $B_2=-2.3 \cdot 10^{11}$ nm$^3$ and $-2.4 \cdot 10^{11}$ nm$^3$ for a depletant concentration $\phi_{d}=0.20$ and a square-well attraction of $\varepsilon = -7 k_{\mathrm{B}}T$ respectively.
Therefore, we can approximate the effective pair interaction with a square-well attraction in simulations, even though the shape of the potentials is quite different.

Based on the size of the roughness ($r\approx$ 70~nm) and the much smaller radius of gyration of the depletant, only a negligible attraction between the rough lobes of the particles is expected\cite{Kraft2012}. However, at depletant volume fractions above $\phi_{d} = 0.33$, aspecific interactions (rough-smooth and rough-rough) became prevalent; at such high concentrations of depletant even the small overlap volume between rough surfaces was sufficient to cause a net attraction. Therefore, the interactions between MM particles can be considered specific up to $\phi_{d} = 0.33$.  Since this depletant concentration corresponds to $\varepsilon \approx -14\;k_{\text{B}}T$, the ears of the particles can indeed safely be considered as hard objects throughout the whole range of the simulations form $\varepsilon = -4\;k_{\text{B}}T$ to $-12\;k_{\text{B}}T]$.

\subsubsection{Particle Aggregation}

In general, cluster formation in the capillaries was found to stabilize in 2-3 days. Hardly any change in cluster composition was observed during subsequent days. Particles and clusters were found homogeneously suspended throughout the capillary, indicating that gravity played no part in the formation of these structures. Also, the density matching with D$_{2}$O proved sufficient to prevent sedimentation while the samples were stationary for microscope observation. In samples with a low particle volume fraction, $\phi_{\mathrm{particles}} = 0.003$, cluster formation was only observed for $\phi_{d} > 0.32\;\rho_{overlap}$, corresponding to $\varepsilon \approx -13\;k_{\text{B}}T$. At this particle volume fraction, primarily small clusters were formed. By letting the clusters sediment to the glass wall prior to observation, the orientation of the individual particles can be observed reasonably well for clusters of a few particles and their structure can be investigated using optical microscopy.
\begin{figure}[h]
\centering
        \includegraphics[width=1.0\textwidth]{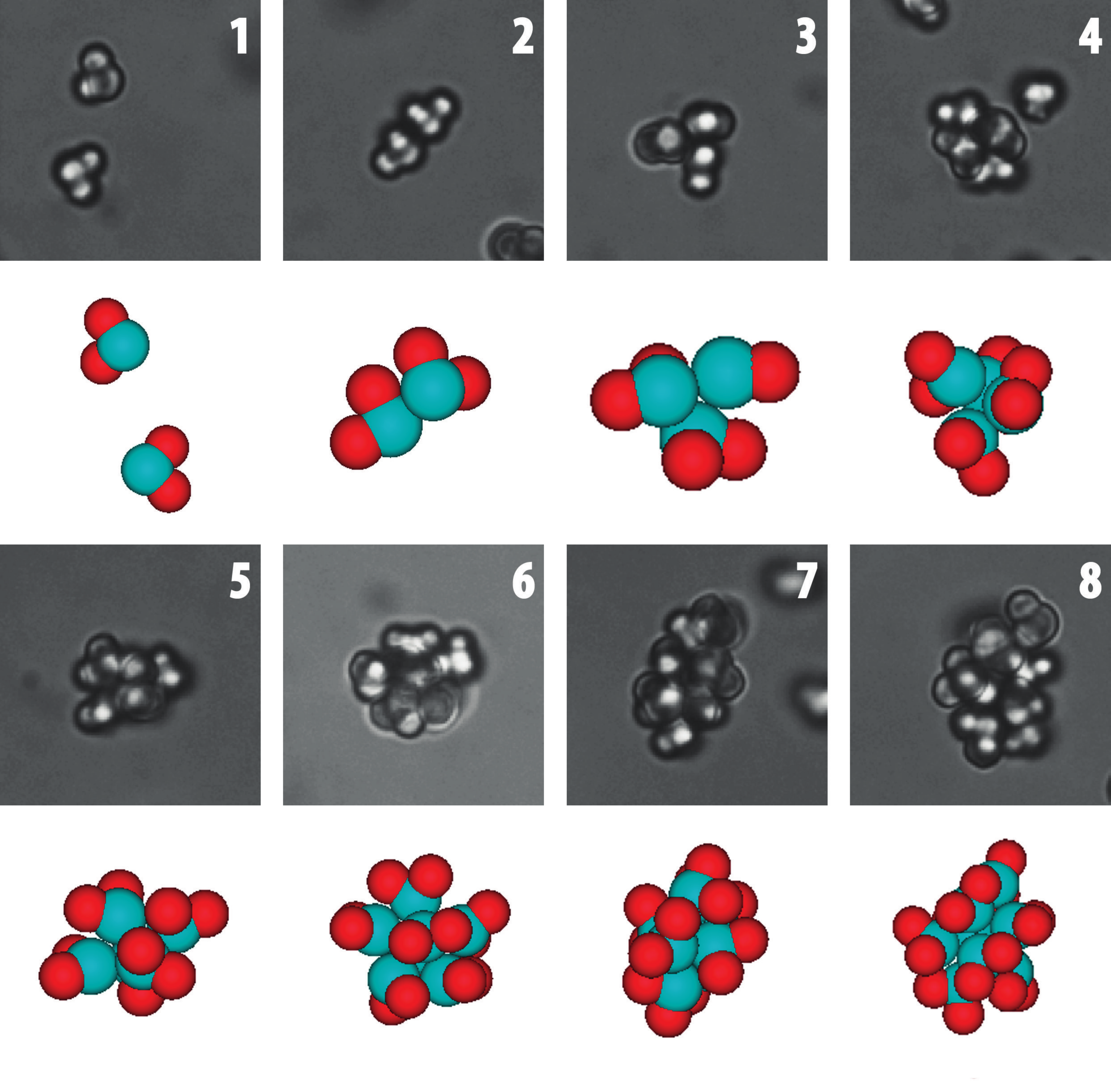}
\caption{\small{Optical micrographs of clusters of different sizes found at low particle concentration $\phi_{\mathrm{particles}} = 0.003$ compared to similar clusters found in simulations with the smooth and rough sides colored blue and red, respectively.
\label{fig:specificinteractions}}}
\end{figure}
Fig.~\ref{fig:specificinteractions} shows optical micrographs of typical clusters of 1 to 8 particles next to similarly shaped clusters found in simulations, clarifying that the particles indeed specifically stick with their smooth sides together. The figure also shows that as clusters get larger, the positions and orientations of the individual particles become harder to observe even when completely sedimented. It would nevertheless be interesting to investigate how the cluster structure develops as clusters increase in size. In computer simulations, we observed that the small clusters can grow out into longer, well-defined structures at any concentration. However, in the experiments at $\phi_{\mathrm{particles}} = 0.003$, cluster growth seemed inhibited due to the rapid depletion of free particles. Also, recombination of existing clusters into tubes was likely difficult because of the limited diffusion of the larger clusters. While increasing the depletant concentration further was found to induce aspecific interactions, increasing the total particle concentration was found to induce the formation of longer structures, which is described in the next section.

\subsubsection{Growth of Tubes and Structure Morphology}

\begin{figure}[h]
\centering

        \includegraphics[width=1.0\textwidth]{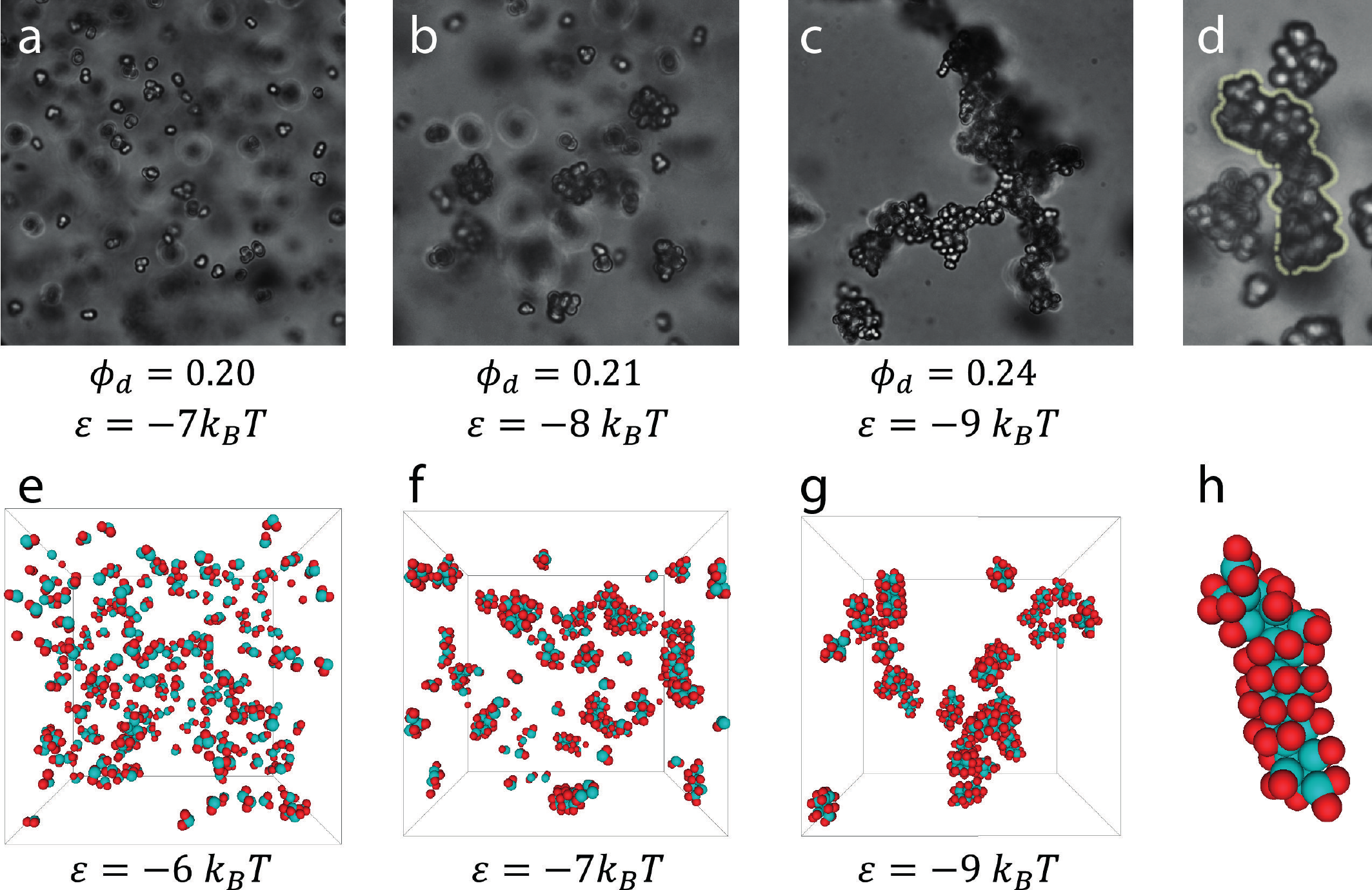}

\caption{\small{(a to c:) Optical micrographs of structures found in samples with volume fraction $\phi_{\mathrm{particles}} = 0.01$ for a depletant volume fraction $\phi_{d}$ of 0.20 to 0.24, corresponding to an attractive interaction potential $\varepsilon$ of $-7\;k_{\text{B}}T$ to $-9\;k_{\text{B}}T$. (e to g:) The MC-simulations at $\phi_{\mathrm{particles}} = 0.01$ show the same resulting structures at comparable values of $\varepsilon$. (d and h:) Close-ups of comparable tube-like clusters from experiments (highlighted) and simulations, respectively. 
\label{fig:structures}}}
\end{figure}

To observe the formation of larger self-assembled structures of MM particles, the volume fraction $\phi_{\mathrm{particles}}$ in the experiments was increased from $0.003$ to $0.01$. It was found that at this higher volume fraction, a lower depletant concentration, indicating a lower attractive interaction potential, was required to form clusters, and that larger structures were formed. For comparison, optical micrographs from the experiments and snapshots from computer simulations at $\phi_{\mathrm{particles}}=0.01$ are shown in Fig.~\ref{fig:structures} for different attraction strengths.

Upon increasing the attractions, a trend can be observed from first small clusters and free single particles (Fig.~\ref{fig:structures}.a and e), to larger clusters (Fig.~\ref{fig:structures}.b and f), and finally tube-like structures (Fig.~\ref{fig:structures}.c and g). The larger clusters seem to grow in only one dimension, resulting in tube-like structures with a well-defined diameter that grow seemingly unbounded in length and in random directions. A close-up of a single representative tube-like cluster from experiments and simulations is provided in Fig.~\ref{fig:structures}.d and h respectively. In the experiments, structures were found to branch occasionally (Fig.~\ref{fig:structures}.c).

The shape of these structures can be attributed to the properties of the building blocks. While the attractive lobes of the particles attract each other and form the backbone of the tube-like structures, the rough lobes act as a steric constraint, preventing bond formation and structure growth on two sides of each particle at an angle of 90\textdegree. This means that if 3 or 4 building blocks are in a planar configuration, new particles can only attach from the two out of plane directions, i.e. either from below or above, promoting the growth of tubular-shaped aggregates.

From the final configurations in the computer simulations, we analyzed the structure of the tube-like clusters in more detail. We found that the particles can be stacked in two different ways, resulting either in straight or spiral-like tube structures. Both of these structures 
occur next to each other in the simulation box, and we even found tubes that were partially made up of one structure and partially of the other one. 

\begin{figure}[]
\centering
\includegraphics[width=1.0\textwidth]{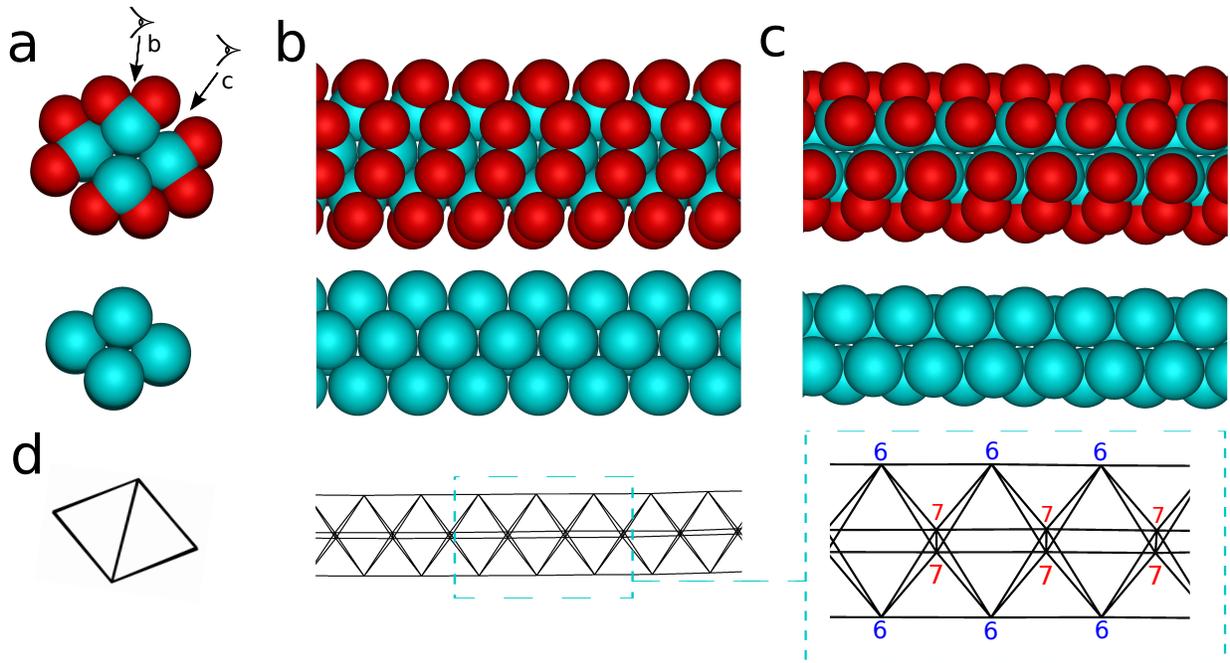}
\caption{\small{Straight tubular structure with the lowest binding energy found in computer simulations. The smooth lobes are shown in blue, the steric ears are shown in red. (a:) top-view showing the Mickey Mouse particles (b-c:) side-views as seen from the directions denoted in (a) (d:) representation of the structure showing the bonds between the MM particles as sticks, for the same viewing angles as in (a) and (b). The mickey mouse particles are located at the vertexes. The numbers denote the number of connections for each vertex. \label{fig:groundstate}}}
\end{figure}

An example of a straight tube-like structure is shown in Fig.~\ref{fig:groundstate}a-c. For each viewing angle, the entire MM particles or only the smooth lobes are shown. The smooth lobes on the inside form a closely packed backbone, whereas the steric ears are all located on the outside and shield the backbone, thus preventing branching of the structure. In Fig.~\ref{fig:groundstate}(d) the bonds between the smooth spheres of the MM particles are depicted as sticks. It can be seen that the bonds form a framework of octahedrons with tetrahedrons in between. The MM particles located at the vertexes of the square of the octahedron form 7 bonds with surrounding particles, whereas the particles on the top ends of the octahedrons are connected with 6 bonds, making up for an average of $6.5$ bonds per particle. In addition to the NVT simulations, we performed Monte Carlo simulations in which a small number of MM particles was compressed in a rectangular box whose 3 lattice vectors were free to change independently. From this, it appeared that the straight tube structures is the ground state tube-structure with the maximum number of bonds per particle.

\begin{figure}[]
\centering
\includegraphics[width=1.0\textwidth]{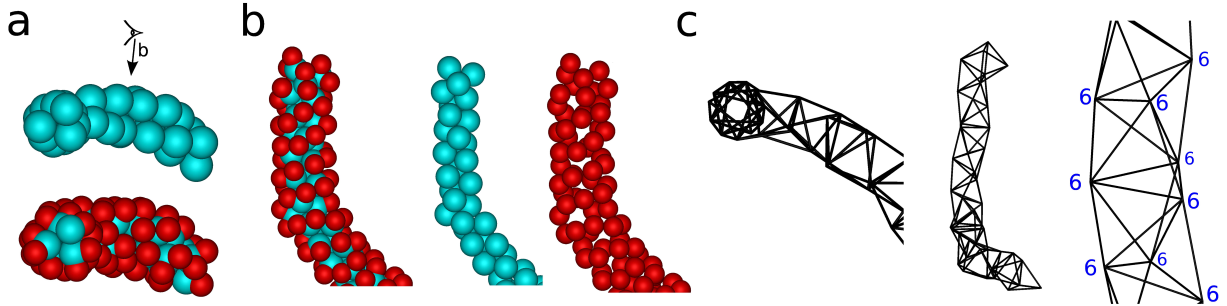}
\caption{\small{Tubes with the structure of a Bernal spiral found in computer simulations. The smooth lobes are shown in blue, the steric ears are shown in red. (a:) top-view showing the Mickey Mouse particles (b:) side-views as seen from the directions denoted in (a) (c:) representation of the structure showing the bonds between the MM particles as sticks, for the same viewing angles as in (a) and (b). The mickey mouse particles are located at the vertexes. The numbers denote the number of connections for each vertex. \label{fig:spiral}}}
\end{figure}

In addition to straight tubes, the MM particles can also be stacked in a spiral-like fashion, as is shown in Fig.~\ref{fig:spiral}. In Fig.~\ref{fig:spiral}a-b, top and side views of a spiral observed in computer simulations are shown. The framework of bonds making up the structure is shown in Fig.~\ref{fig:spiral}c, revealing a repeated structure of connected face-sharing tetrahedra also known as a Bernal spiral~\cite{bernal1964bakerian} or Boerdijk-Coxeter helix~\cite{Boerdijk,Coxeter}. In a perfect spiral, each particle is bound to six neighbours. Bernal spirals were already found for colloidal systems with short-ranged attractions and screened electrostatic repulsion~\cite{campbell2005dynamical, sciortino2005one} but recently also for Janus particles that are able to undergo directional interactions~\cite{chen2011supracolloidal}. Recent efforts have focused on creating Bernal spirals by design with patchy particles~\cite{fejer2014design, morgan2013designing}. However, in either case the potential or the positions, size and shape of the patches needs to be carefully tuned. In our case, the spiral forms mainly for geometric reasons, illustrating that the steric ears on the Mickey Mouse particles constitute a robust way of creating functional directional interactions. In addition, the ears also protect the attractive backbone of the spiral from sideways attachments and thus provides a new possible way to produce core-shell tubular structures by self-assembly.

\begin{figure}[h]
\centering
\includegraphics[width=0.5\textwidth]{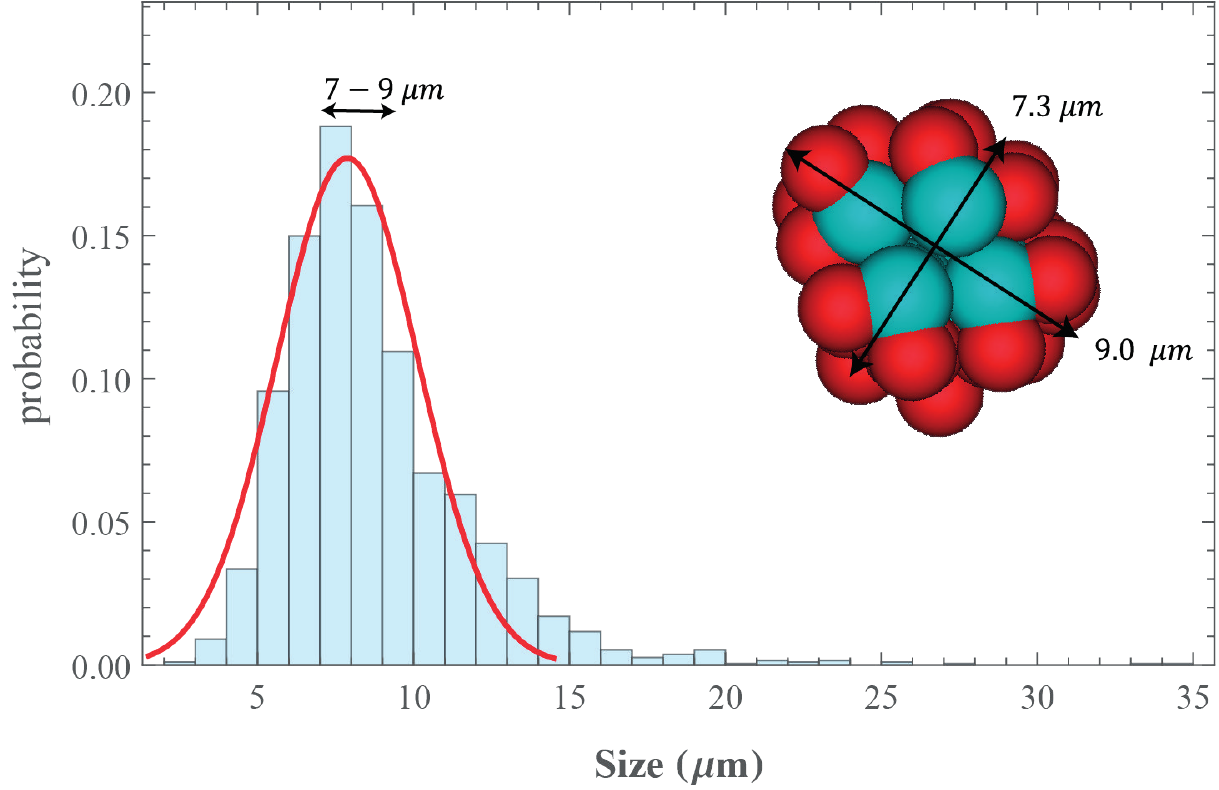}
\caption{\small{Distribution of tube diameters obtained from optical microscopy, with a Gaussian fit (in red). The diameter is well-defined around 7 to 9~\textmu m. The insets show a cross-sections of the straight and spiral tube structures found in simulations, the diameters of which correspond to the peak in the distribution.
\label{fig:diameterdistribution}}}
\end{figure}

In the experiments, the 8-particle cluster in Fig.~\ref{fig:specificinteractions} and the clusters in Fig.~\ref{fig:structures} show that for increasing cluster size, the scattering caused by the large refractive index mismatch between polystyrene and water ($\Delta n_D = 0.2592$) makes it more and more difficult to assess the internal structure using optical microscopy. What can be determined, however, is the diameter of the observed tubes. Fig.~\ref{fig:diameterdistribution} shows a narrow distribution of cluster diameters. A Gaussian fit made to this distribution ($\mu = 8\;\text{\textmu m},\; \sigma = 2\;\text{\textmu m}$) corresponds well to the diameters of the tubular structures (7 to 9~\textmu m) found in the simulations (Fig.~\ref{fig:structures}.h, Fig.~\ref{fig:groundstate} and Fig.~\ref{fig:spiral}, see also the insets in Fig.~\ref{fig:diameterdistribution}). The additional spread is caused by the random orientation of the rough lobes, which can make a difference of 1~\textmu m, depending on whether one lobe ($R_r = 1.17$~\textmu m) is oriented perpendicular to the tube or both are turned out of plane.
The tail on the right side of the distribution can be explained by the presence of branch points, which have a larger diameter.

As can be seen in Fig.~\ref{fig:groundstate} and Fig.~\ref{fig:spiral}, the attractive smooth lobes are stacked in an ordered fashion. However, as the smooth lobes lobes are free to rotate, the steric ears can also change position and essentially form a concentrated quasi-2D system of hard dimers on a cylinder. This gives rise to an inherently disordered, but heavily correlated configuration of rough lobes that dictates the direction from which additional particles can attach to the cluster and thereby contributes to the cluster structure.

This mechanism can also cause branching of the tubes by occasionally allowing sufficient space between the rough lobes for a new particle to attach sideways (Fig.~\ref{fig:structures}.c). However, this branching was only observed in experiments and not in simulations. Additionally, while the structure of the experimental clusters have the same diameter as the simulated tubes, they generally do not extend in one direction long enough to form a spiral (Fig.~\ref{fig:structures}.c and d).
It is therefore likely that also undesired experimental factors like particle size polydispersity and occasional rough-smooth attractive contacts contribute to disorder and branching.

\begin{figure}
            \includegraphics[scale=0.5]{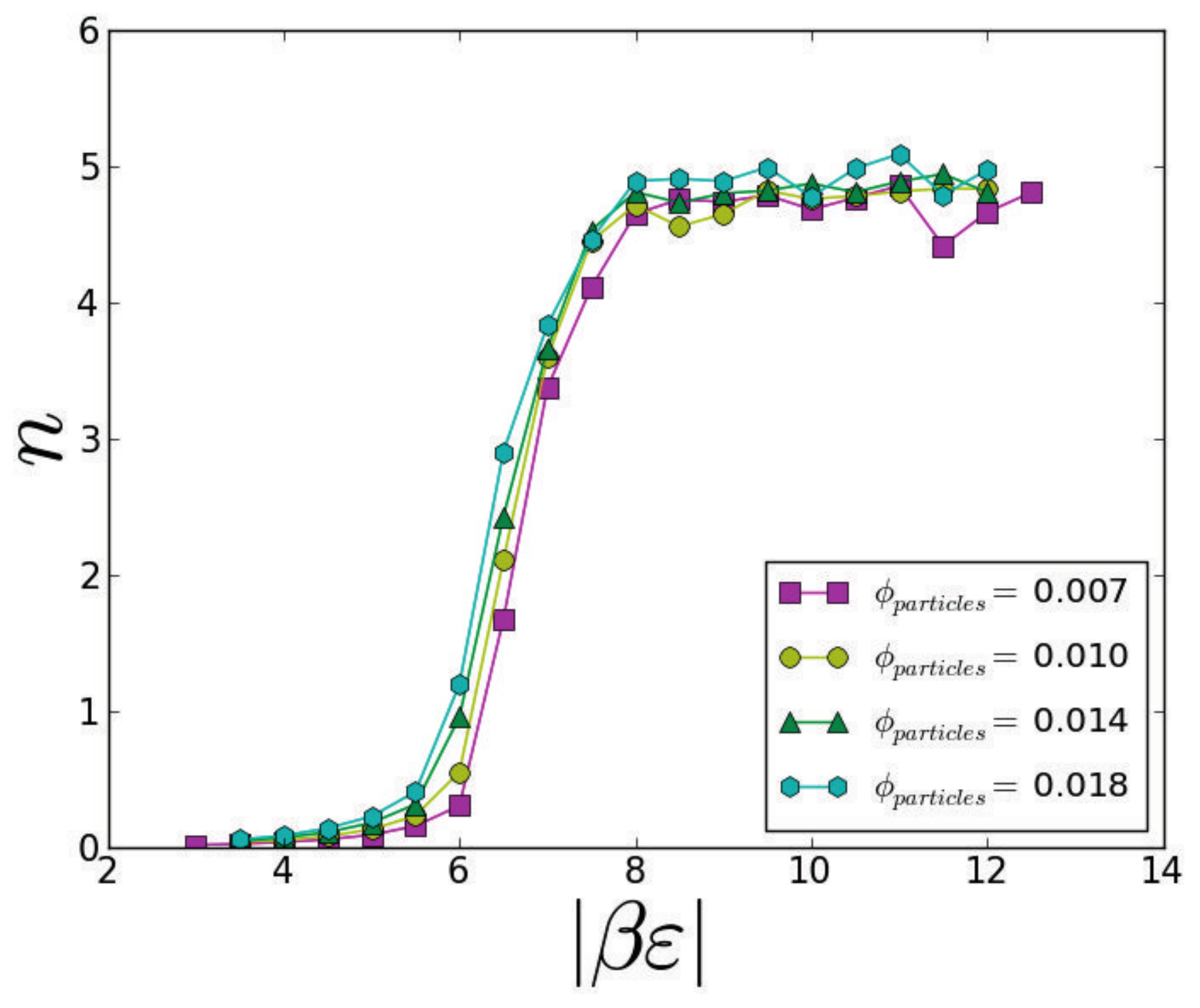}
            \caption{\small{Average number of contacts per particle $\langle n \rangle$ in a cluster as function of the well depth of the attractive pair interaction, for different volume fractions $\phi_{\mathrm{particles}}$. The figure shows a sharp transition from only small clusters at low interaction strength to a constant number of contacts ($\sim$5) for stronger interactions, irrespective of $\phi_{\mathrm{particles}}$. }} 
            \label{fig:contVsEps}
\end{figure}

From the computer simulations, we obtained the average number of contacts per particle $\langle n \rangle$ for a range of attraction strengths $\epsilon/k_{\mathrm{B}}T$. In Fig.~\ref{fig:contVsEps}, $\langle n \rangle$ is plotted against $\varepsilon$ for different values of $\phi_{\mathrm{particles}}$. Fig.~\ref{fig:contVsEps} shows a sharp transition from only small clusters with low $\langle n \rangle$ at low values of $\lvert \varepsilon \rvert$, to a constant value of $\langle n \rangle \approx 5$ at higher interaction strength. This sudden jump in the internal energy of the system was also found for a tube-forming system of one-patch particles and might indicate a degree of cooperative polymerization \cite{coop}. The interaction strength $\varepsilon$ at which the formation of the tubes started did not appear to be sensitive to the fraction $\phi_{\mathrm{particles}}$ for the region of volume fractions investigated. 

In the case all particles had been incorporated in tube-like structures, the average number of bonds would have been expected to converge towards $\langle n \rangle \sim 6-6.5$. However, we have observed that in the computer simulations for attractions of $\varepsilon < -7\;k_{\text{B}}T$, the number of bonds had not yet converged to this number, indicating that the simulations were not fully equilibrated. To check if the same holds for the experiments, the timescales at which bonds between particles can break were calculated from the interaction potential using Kramers' approach~\cite{Kramers1940} in the same way as performed previously~\cite{Kraft2012}. In our system, the escape time for a single bond was found to range from 10~s for $\varepsilon = -7\;k_{\text{B}}T$ to 10 minutes for $\varepsilon = -12\;k_{\text{B}}T$. Escape times on the order of seconds for a single bond mean that pairs can break and form freely, and also small clusters with up to 3 bonds per particle, which can be broken sequentially, can rearrange their structure. However, once a particle has formed 4 or more bonds, it is no longer possible to remove the particle without breaking at least 2 bonds simultaneously. Escape times for such configurations quickly rise to $10^{5}$~s for breaking 2 bonds simultaneously and $10^{12}$~s for breaking 4 bonds, far beyond experimentally observable time scales. This means that once tube-like structures with 4 or 5 sufficiently strong bonds per particle are formed, the system gets trapped in this state and can no longer optimize its structure by breaking and reforming bonds. Although the internal structure of the clusters found in both experiments and simulations appears to be kinetically trapped, some particles can still exchange with the free particles in the medium, establishing a (local) equilibrium between the tubes and free particles.

\subsubsection{Cluster structure and free particle concentration}
Fig.~\ref{fig:structures} illustrates that upon increasing the interaction strength, the transition from hardly any clusters to an almost completely clustered system goes through a regime where both clusters and free particles are present. This concentration of free particles $\phi_{\mathrm{free}}$ is easily determined in both simulations and experiments and can be related to the binding free energy of a particle to provide estimate information on the experimental cluster structure such as the average number of bonds.

The concentration of particles at which clusters start forming can be regarded as a critical micelle concentration (\emph{cmc}) of a system of surfactants and consequently, just like with surfactants, this \emph{cmc} is also the concentration of free particles present in the system once clusters are formed. This concentration can be related to the free energy difference between free particles and particles bound in a cluster via\cite{Kraft2012}
\begin{equation}
\phi_{\mathrm{free}} = \phi_{\mathrm{cmc}} \simeq \frac{v_{\mathrm{MM}}}{v_{av}} e^{\Delta u/k_{\text{B}}T}
\label{eq:phivsu}
\end{equation}
where $\phi_{\mathrm{free}}$ is the volume fraction of free particles, $v_{\mathrm{MM}}$ is the volume of a colloid, $v_{av}$ is a measure for the available free volume of a particle in a cluster and $\Delta u = \frac{\mu_{n}^{\circ}-n\mu_{1}^{\circ}}{n-1}$, where $\mu_{n}^{\circ}$ is the chemical potential of a cluster of $n$ particles and $\mu_{1}^{\circ}$ is the chemical potential of a single particle. For large clusters ($n \gg 1$), $\Delta u$ equals the average bonding free energy difference between a free particle and a particle in a cluster. Due to the system's short interaction range, bonds can be considered pairwise additive and the energy difference per particle is just half the number of bonds times the pair energy: $\Delta u =\frac{\langle n \rangle}{2}\varepsilon$. The $v_{av}$ term in Eq.~\ref{eq:phivsu} is a measure of the translational entropy still available to particles in a cluster. The difference in rotational entropy between free particles and particles in a cluster is not taken into account in this equation, since particles bound by depletion interaction via their smooth lobes only lose a fraction of their rotational freedom due to geometric constraints.

The majority of particles in the tube-like clusters have the same number of bonds and, related to that, a similar value of $v_{av}$. A particle inside a cluster is confined in 3 dimensions such that $v_{av}=\xi^3$, where $\xi$ is the width of the potential well the particle is in. This means that if there is an equilibrium between free particles and tube-like clusters in the system, the logarithm of the concentration of free particles, $\ln\phi_{\mathrm{free}}$, should depend on $\varepsilon$ via:
\begin{equation}
\ln\phi_{\mathrm{free}} = \ln(\frac{v_{\mathrm{MM}}}{\xi^3})+ \frac{\langle n \rangle}{2} \varepsilon
\label{eq:lnPhiVsEps}
\end{equation}

\begin{figure}[h]
\centering
\includegraphics[width=0.8\textwidth]{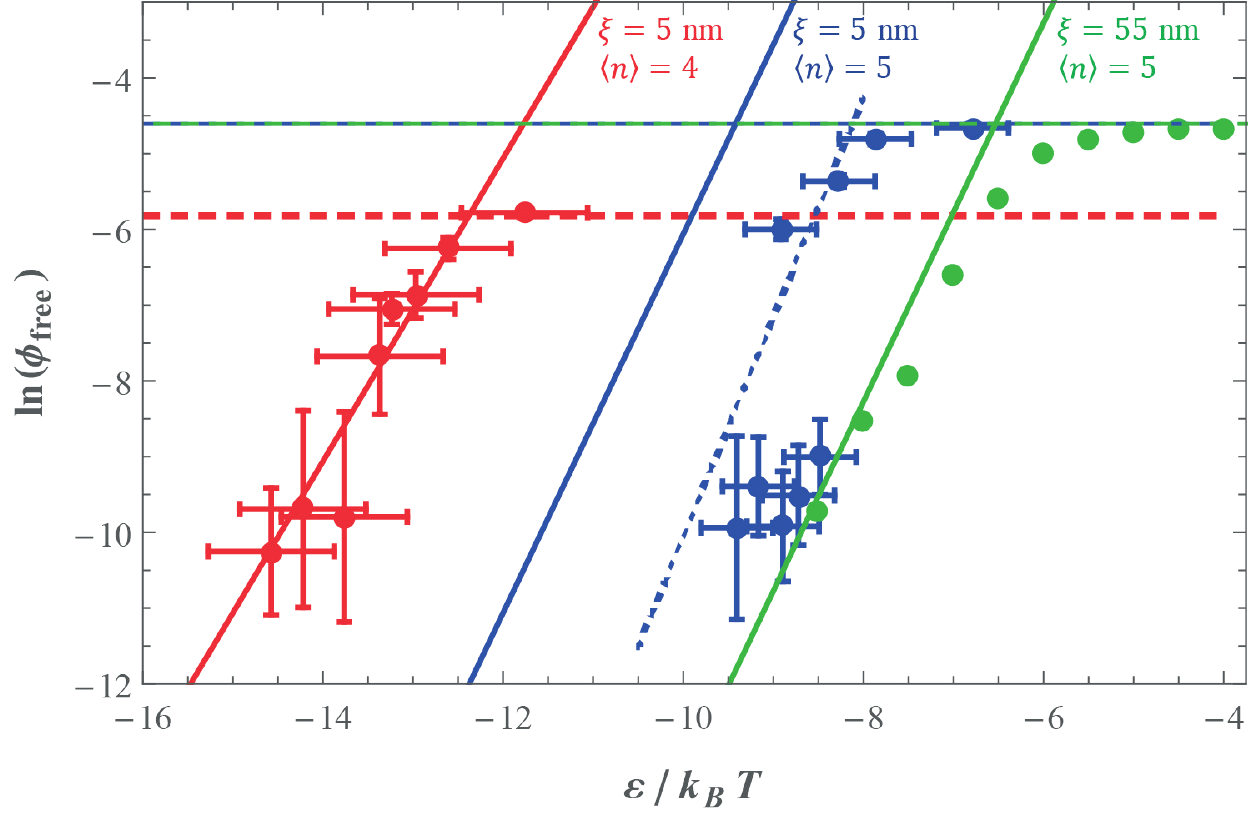}
\caption{\small{$\ln\phi_{\mathrm{free}}$ as function of the system's pair interaction found for experiments at volume fractions $\phi_{\mathrm{particles}}=0.003$ (red) and $\phi_{\mathrm{particles}}=0.01$ (blue) and for computer simulations ($\phi_{\mathrm{particles}}=0.01$, green) plotted against $\varepsilon$. For the experiments, $\varepsilon$ was calculated from experimental conditions (by finding the minimum of Eq.~\ref{eq:netinteraction}). For the simulation data, the depth of the square well $\varepsilon$ was used. The solid lines represent plots of Eq.~\ref{eq:lnPhiVsEps}, using the values shown at the top as input parameters. The dashed horizontal lines mark the total particle concentrations $\phi_{\mathrm{particles}}$. The dotted blue line is a fit of Eq.~\ref{eq:lnPhiVsEps} finding $\langle n \rangle = 5.8$. \label{fig:lnphivsekT}}}
\end{figure}

Fig.~\ref{fig:lnphivsekT} shows the logarithm of the free particle concentration $\ln\phi_{\mathrm{free}}$ for the simulations at $\phi_{\mathrm{particles}}=0.01$ and the experimental systems with the lower ($\phi_{\mathrm{particles}}=0.003$) and higher ($\phi_{\mathrm{particles}}=0.01$) particle concentration. At low interaction strength, $\varepsilon/k_{\mathrm{B}}T \sim 0$, only free particles are expected as in this case the total particle concentration is lower than the critical micelle concentration, $\phi_{\mathrm{particles}} < \phi_{\mathrm{cmc}}$. At sufficiently strong attractions, tubes and free particles are present and according to Eq.~\ref{eq:lnPhiVsEps}, $\ln\phi_{\mathrm{free}}$ should depend linearly on $\varepsilon$, as is shown by the solid lines assuming values for $\xi$ obtained from the shape of the interaction potential and a slope dependent on $\langle n \rangle$. At higher interaction strengths, from $\varepsilon \approx -8.5\;k_{\mathrm{B}}T$ in the simulations, almost all particles are part of a cluster and the fraction of free particles becomes too low to detect.  Also in the experiments, hardly any free particles are found at higher values of $\lvert \varepsilon \rvert$, resulting in low concentrations ($\ln\phi_{\mathrm{free}} \approx -10$) with large error bars.

In the computer simulations performed for $\phi =0.01$, upon increasing the attraction strength, the logarithm of the concentration of free particles was found to remain roughly constant up to $\varepsilon \approx -6\;k_{\text{B}}T$, followed by a linear decrease over a range of $~2\;k_{\text{B}}T$. For the simulations,  $\xi$ is simply the width of the square well ($0.02 \sigma_s = 55$~nm) and $\langle n \rangle \approx 5$ (see Fig.~\ref{fig:contVsEps}). Using these numbers, the measured $\phi_{\mathrm{free}}$ from the simulations is in good agreement with Eq.~\ref{eq:lnPhiVsEps} in the region where tubes form.

The experiments performed for volume fractions $\phi_{\mathrm{particles}}=0.003$ and $\phi_{\mathrm{particles}}=0.01$, also show a decrease in $\ln \phi_{\mathrm{free}}$ upon increasing attraction strength, from a point where all the particles are free, so $\phi_{\mathrm{particles}} \approx \phi_{\mathrm{free}}$, to a point where almost all particles are part of a larger cluster and $\phi_{\mathrm{free}} << \phi_{\mathrm{particles}}$. In the experiments, $v_{av}$ is the volume that a bound particle can explore using its thermal energy, which can be approximated by the width of the potential well (Fig.~\ref{fig:pairpotential}) $1\;k_{\text{B}}T$ above the potential minimum (approximately $5$~nm in this case). 
We found that for the experiments at $\phi_{\mathrm{particles}}=0.003$, the data points show a good agreement with Eq.~\ref{eq:lnPhiVsEps} using $\langle n \rangle = 4$. The small clusters in this system, like the ones shown in Fig.~\ref{fig:specificinteractions}, suggest that this is a reasonable estimate for the number of bonds in the experiments at this concentration.

At a higher volume fraction  $\phi_{\mathrm{particles}}=0.01$, where tube-like structures were observed also in experiments, the decrease of $\ln \phi_{\mathrm{free}}$ with increasing attraction strength lies closer to the simulation values of the same $\phi_{\mathrm{particles}}$ in Fig.~\ref{fig:lnphivsekT}. Because the tube-like structures in experiments at $\phi_{\mathrm{particles}}=0.01$ are comparable to those in the simulations in appearance (Fig.~\ref{fig:structures}.d and h) and dimensions (Fig.~\ref{fig:diameterdistribution}), one would expect that $\langle n \rangle \approx 5$, which indeed provides reasonable agreement with the experimental data. However, we find a better fit to the measured data for $\langle n \rangle \approx 5.8$ (The dotted line in Fig.~\ref{fig:lnphivsekT}), indicating that the number of contacts is probably higher. This might be due to the the presence of longer clusters with more branch points in the experiments compared to simulations.

The free particle concentration from both the simulations and experiments are in good agreement with Eq.~\ref{eq:lnPhiVsEps}, even though the system is not fully equilibrated. The trend in $\ln \phi_{\mathrm{free}}$ provides a reasonable estimate of the number of bonds based on $v_{av}$ and $\varepsilon$.
The deviation from linearity that is present in the experiments, especially for higher depletant concentrations at $\phi_{particles}=0.01$, can be attributed to the uncertainty in determining $\varepsilon$ from the experiments. 
Firstly, as mentioned before, the surface potential of the smooth side of the particle is unknown and the value used to determine $\varepsilon$ is most likely an overestimation. Secondly, the experimental error in determining $\varepsilon$, about $0.5\;k_{\text{B}}T$, covers over a quarter of the total range where the decreasing free particle concentration can be determined.

The number of bonds obtained from the free particle concentration in the experiments corresponds to the overall $\langle n \rangle$ from simulations that were not fully equilibrated (Fig.~\ref{fig:contVsEps}). Both values are lower than the $6$ or $6.5$ contacts for fully developed tubes (see Fig.~\ref{fig:groundstate} and Fig.~\ref{fig:spiral}), confirming that the tube-like clusters do not converge to their ground state under the present conditions.

\section{\label{sec:summary}Conclusions}
             
In this paper, we have reported the synthesis and structure formation of anisotropic ``Mickey Mouse'' (MM) particles, consisting of one central smooth lobe connected to two rough lobes on either side at an angle of $\sim$90\textdegree . By tuning the amount of added depletant, attractions can be induced between the smooth lobes, whereas the rough lobes remain repulsive. The attractions were shown to occur exclusively between the smooth lobes of the particles over a considerable range of depletant concentrations. The size of our particles enables observation using optical microscopy while successfully suppressing the effect of gravity by dispersing them in a mixture of H$_{2}$O and D$_{2}$O and tumbling the samples between measurements. 

In the experiments, at sufficiently strong attractions and sufficiently high particle concentration, the MM particles were shown to form tube-like structures. This is different from the spherical micelle-like clusters observed before for colloids with a single lobe~\cite{Kraft2012}. By simulating the MM particles represented by one attractive and two repulsive spheres, we were able to complement the information on internal structure missing from the microscopy experiments. Both in simulations and experiments, these particles were found to self-assemble into tube-like structures of seemingly unbounded length, but with a well-defined diameter. The structure of the tubes in the simulations consisted of a mixture of straight fragments and fragments in the shape of Bernal spirals. It is however hard to access equilibrium states in this experimental system, as the short ranged interactions and multiple bonds that can be formed by each colloid prevent them from effectively exploring many configurations on the experimental timescale. Although the predominant structure in the experiments therefore appears to be kinetically trapped, we could use 
the bond strength and bonding volume of the particles to get reasonable estimates for the number of bonds in the tube-like structures from concentration of free particles in the system.

Finite-sized clusters have been found experimentally for Janus particles with one sticky patch~\cite{Chen2011d} and patchy dumbbells with an attractive lobe~\cite{Kraft2012} and are supported by computer simulations~\cite{Chen2007a, Sciortino2009}. This work shows that the shape of the building block can be tuned to influence the final structure that is formed. In this case, the introduction of an additional non-sticky lobe even completely changed the dimensionality of the formed structures, marking important new steps on the road to the experimental realization of complex self-assembled colloidal structures. In particular, we highlight the formation of Bernal spirals of Mickey Mouse particles, as a direct result of the combination of smooth lobe providing a selective short-range depletion attraction and steric ears facilitating directionality through an effectively long-ranged shielding. This is especially interesting as in previous works these structures were established either through careful tuning of the interaction potential~\cite{campbell2005dynamical, sciortino2005one} or by well-defined attractive patches on the surface of spherical particles~\cite{chen2011supracolloidal, fejer2014design, morgan2013designing}. Since most patchy systems are, like the systems studied here, based on short-ranged interactions and considering the drive towards more complex patchy colloids, the results presented in this study are expected to play an increasingly important role in the practical realization of colloidal self-assembly.

\bibliography{mickeyrefs}

\end{document}